\documentclass[emulateapj5]{aastex}

\usepackage{emulateapj5}
\usepackage{psfig}

\def\arcsec{$\,^{\prime\prime}$~}
\def\arcmin{$\,^\prime$~}

\def\erg/cm2sec{ergs~cm$^{-2}$~s$^{-1}$}  
\def\ergcm2{ergs~cm$^{-2}$}  
  
\def\mdot{$\dot{m}$~}

\newcommand{\lsim }{{\lower0.8ex\hbox{$\buildrel <\over\sim$}}}
\newcommand{\gsim }{{\lower0.8ex\hbox{$\buildrel >\over\sim$}}}

\def\aap{ A\&A}

\def\simge{\mathrel{%
   \rlap{\raise 0.511ex \hbox{$>$}}{\lower 0.511ex \hbox{$\sim$}}}}
\def\simle{\mathrel{
   \rlap{\raise 0.511ex \hbox{$<$}}{\lower 0.511ex \hbox{$\sim$}}}}

\newcommand{\Msun}{\ifmmode {M_{\odot}}\else${M_{\odot}}$\fi}
\newcommand{\Rsun}{\ifmmode {R_{\odot}}\else${R_{\odot}}$\fi}

\newcommand\HST{{\it HST}}
\newcommand\Chandra{{\it Chandra}}

\shorttitle{Two Neutron Stars in 47 Tucanae}
\shortauthors{Heinke et al.}

\begin{document}
\title{X-ray Studies of Two Neutron Stars in 47 Tucanae: Toward
Constraints on the Equation of State}  

\author{C. O. Heinke, J. E. Grindlay, D. A. Lloyd, P. D. Edmonds}
\affil{Harvard-Smithsonian Center for Astrophysics \\ 
60 Garden Street, Cambridge, MA  02138\\ cheinke@cfa.harvard.edu}

\begin{abstract}
We report spectral and variability analysis of two 
quiescent low mass X-ray binaries (X5 and X7, previously detected with
the ROSAT HRI) in a 
{\it Chandra} ACIS-I observation of the globular cluster 47 Tuc.  X5
demonstrates sharp eclipses with an 
8.666$\pm0.01$ hr period, as well as dips 
showing an increased $N_H$ column.  The thermal spectra of X5 and X7 are
well-modeled by unmagnetized hydrogen atmospheres of hot neutron
stars.  No hard power law component is required. A possible edge or absorption
feature is identified near 0.64 keV, perhaps an OV edge from a hot
wind.   Spectral fits imply
that X7 is significantly more massive than the canonical 1.4 \Msun\
neutron star 
mass, with M$>1.8$\Msun\ for a radius range of 9-14 km, while X5's
 spectrum is consistent with a neutron star of mass 1.4$M_{\odot}$ for
the same radius range.  Alternatively, if much of the X-ray luminosity
is due to continuing 
accretion onto the neutron star surface, the feature may be the 0.87
keV rest-frame absorption complex (O VIII \& other metal lines)
intrinsic to the neutron star atmosphere, and a mass of 1.4 \Msun\ for X7 may
be allowed.  

\end{abstract}

\keywords{
accretion disks ---
binaries: close, eclipsing ---
binaries: X-ray ---
globular clusters: individual (NGC 104) ---
stars: neutron 
}

\maketitle

\section{Introduction}

The primary scientific goal of neutron star (NS) mass and radius
measurements are to constrain 
the NS equation of state (EOS) and thus internal composition (e.g.,
neutrons and protons vs. hyperons or free quarks). The NS EOS would
 be sharply constrained by a firm determination 
of a NS mass above 1.6 \Msun.  This would eliminate all EOSs with
significant softening (van Kerkwijk 2001). Bose-Einstein
condensation of kaons, hyperons, and pions in the core would greatly
soften the equation of state, leading to NSs which reach a maximum mass of 
1.4-1.5 \Msun\ (Lattimer \& Prakash 2001).  
  Mass determinations of high accuracy have been made for twelve neutron
stars in binary systems using general relativistic effects, ten of
them in double NS systems.  Thorsett \& Chakrabarty (1999) find that a
Gaussian distribution with mean=1.35\Msun\ and $\sigma$=0.04\Msun\ well
describes the masses of these NSs, with none above 1.45 \Msun.
However, at least two NSs in accreting X-ray systems show evidence for higher
masses.  Vela X-1 has a reported mass of $1.86\pm0.32$\Msun\ (95\%
confidence), though systematic deviations in radial velocity
determinations continue to make this result uncertain (Barziv et
al. 2001 and refs therein).  Cygnus X-2 has a reported mass of
$1.78\pm0.23$\Msun\ ($1\sigma$), but this result depends upon the 
modeling of the secondary and disk light curves (Orosz and Kuulkers
1999).  

Other lines of evidence also constrain NS equations
 of state.  For instance, cooling rates of middle-aged
 radio pulsars imply a small
 range of masses around 1.4 \Msun\ and a somewhat stiff EOS (Kaminker
 et al.\ 2001).  Quasi-periodic 
 oscillations in some low-mass X-ray binaries have sometimes been
 interpreted (using various models) to indicate masses near 2 \Msun\
 (see van der Klis 2000 for a review),
 and to force a radius less than 15.2 km (van Straaten et al.\ 2000).
 Also, calculation of the Crab pulsar's moment of inertia (Bejger \&
 Haensel 2002) and modeling of X-ray burst oscillations (Nath et
 al. 2002) suggest that only stiff EOSs are plausible.
 Recent attempts to fit the X-ray and optical
 spectrum of the isolated neutron star RX J185635-3754 seem to have ruled out
 all physically well-understood atmospheres, so mass and radius
 calculations  
 for RX J185635-3754 remain somewhat tentative (see Burwitz et
 al. 2002 and references therein).   

Temperature anisotropies and
 relatively high ($B>10^{11}G$) surface magnetic fields are 
 general features of model predictions for pulsar surface emission, so
 transiently accreting neutron stars (with much lower magnetic fields,
 and negligible temperature anisotropies) may provide a
 simpler path to determining the NS mass and radius if the thermal
 component can be cleanly fit.
Several NSs that have been observed in outburst as soft X-ray
transients have also been detected in quiescence (0.5-2.5 keV X-ray
 luminosity, $L_X,\sim10^{32-34}$
ergs s$^{-1}$), for example Cen X-4 and Aquila X-1 (van Paradijs et
al. 1987, Verbunt et al. 1994).  See Campana et
al.\ (1998a) for a review of soft X-ray transients,
also known as neutron star X-ray novae or quiescent 
low-mass X-ray binaries (qLMXBs). Spectral fits with a soft
(kT=0.2-0.3 keV) blackbody (BB) spectrum, usually 
requiring a hard power-law tail of photon index $1-2$, have been
acceptable but imply an emission area of $\sim1$ km radius, much smaller
than appropriate for a NS surface.  However, heavy elements
settle out of the atmosphere on a 
timescale of seconds (Romani 1987), so pure hydrogen atmospheres
(assuming accretion has occurred from a non-helium secondary) have been
used recently to calculate the expected spectrum. 
Rajagopal and Romani (1996) and Zavlin et al.\ (1996) showed that the
atmosphere of a NS shifts the peak of the emitted radiation to
higher frequencies, due to the strong frequency dependence of
free-free absorption. Thus a blackbody fit will derive a temperature
that is too high and a radius that is too small.   Brown, Bildsten,
and Rutledge (1998, BBR98; see also Campana et al. 1998a) showed that
 the deep crust of a  neutron star transiently accreting with average rate
 $\langle\dot{m}\rangle$ is heated during accretion by 
pycnonuclear reactions, bringing the interior to a steady state
temperature $\sim10^8 \langle$ \mdot$/10^{-10} M_{\sun}$
yr$^{-1}\rangle^{0.4}$ K.  This heating leads to an isotropic  
thermal luminosity between accretion episodes of roughly $L_q=6\times10^{33} 
\langle$ \mdot $\rangle/10^{-10}$ \Msun\ yr$^{-1}$ ergs s$^{-1}$, which
 BBR98 proposed as the source of the luminosity in qLMXBs (the deep
 crustal heating hypothesis).  The power law component in qLMXBs like
 Cen X-4 and Aql X-1, however, has always been attributed to
 continued accretion (see e.g., Campana et al.\ 1998b).

 Fits of unmagnetized ($B\lesssim10^{10}$ G) hydrogen 
atmosphere models to qLMXB spectral data have been consistent with an 
$R_{\infty}$  
($R_{\infty}$ is the effective radius 
seen by a distant observer, $R_{\infty}=R/\sqrt{1-2GM/Rc^2}$,
dependent upon M) of roughly 13 km (Rutledge et
al.\ 1999, 2001a, 2001b). BBR98 noted that the
well-determined distances to globular clusters make them ideal for
determining radii of NSs in qLMXBs by this method.  Recently, Rutledge et
al.\ (2002a) have identified a promising qLMXB candidate lacking a
hard power law 
component in $\omega$ Cen (NGC 5139).  However, this object, and the
soft X-ray transient in NGC 6440 (in't Zand et al. 2001), do not
provide enough photons to constrain mass and radius effectively.  

This picture of ``deep crustal heating'' as being responsible for most
of the emission from quiescent neutron stars is not yet certain,
however.  The alternative is that accretion continues at very low levels
during quiescence, either onto the neutron star surface or onto the
magnetosphere, and that the energy released is directly responsible
for the thermal emission (see Campana et 
al. 1998a).  This continued accretion must not be confined to the
poles to avoid 
producing measurable variability at the NS spin frequency, as tested
by observations of Aql X-1 entering quiescence (Chandler \&
Rutledge 2000).  Although this scenario has not yet produced 
detailed physical models of the spectrum and luminosity, it has received 
support through the observations of short-timescale variability in
Aql X-1 (Rutledge et al. 2002b) and Cen X-4 (Campana et al. 1997),
which cannot be explained by the ``deep crustal heating'' model alone.
The nonthermal power-law tail spectral component also cannot be
explained by deep crustal heating, and therefore suggests continued
accretion.  Accretion onto 
the neutron star surface is thought to give rise to a thermal
spectrum very similar to an equilibrium hydrogen atmosphere,
significantly harder than a blackbody spectrum  
(Zampieri et al. 1995; Deufel, Dullemond, \& Spruit 2001).  If accretion does
indeed provide most of the X-ray luminosity in some qLMXBs, then the
implied mass accretion rates ($\sim2\times10^{-13}$ \Msun
yr$^{-1}$) may be sufficient to maintain some metals in the
photosphere at levels lower than solar (Bildsten et al. 1992;
Brown et al. 1998).
Metals would alter the photospheric opacity, and thus the apparent
radius, as well as introducing possibly observable lines into the
spectrum.  The lack of compelling evidence for such lines in qLMXBs (but
cf. Rutledge et al. 2002b) must be 
considered as evidence against this scenario.

In this paper we  present our spectral and
 temporal analysis of the two brightest 
(L$_X\sim10^{33}$ ergs s$^{-1}$) soft sources in the well-studied
globular cluster 
47 Tucanae (NGC 104).  The deep ROSAT HRI survey of Verbunt \& Hasinger (1998)
resolved 5 sources in the central core, including X5 and X7, but had
 no spectral resolution.  With {\it Chandra}'s
ACIS-I instrument, we have identified 108 sources in the central
2\arcmin$\times$2.5\arcmin region, and are able to conduct spectral
analysis on a few dozen of these sources (Grindlay et al.\ 2001a,
hereafter GHE01).  We show that X5 and X7 have blackbody-like X-ray spectra,
indicating that they are hot NSs.  X5 has been identified  with a
$V=21.6$ counterpart,  
which shows variability and a blue color indicative of a faint accretion
disk (Edmonds et al.\ 2002).  An upper limit for the 
counterpart of X7 is $V\sim23$.  Using
$F_{V}=10^{-0.4V-5.43}$ erg cm$^2$ s$^{-1}$ (V band) gives  $F_X/F_V$
 ratios of 46 and $\gtrsim$183 respectively, which are typical of qLMXB
systems rather than CVs.     In this paper we report
 the detailed spectral analysis of 
X5 and X7, constraining the masses and radii of
these NSs under the assumption of pure hydrogen atmospheres, and 
 identifying a possible signature of low-level accretion.  The large
 numbers of photons (4200 and 5500 from X5 and X7 respectively),
 well-known distance to 47 Tuc, and lack of a significant nonthermal
 component in their spectra allow firmer mass and radius constraints
 (with fewer assumptions) than previous attempts to constrain qLMXB radii.
  
\section{Analysis}

The {\it Chandra} X-ray Observatory observed 47 Tucanae on 2000 March
16-17 for 72 ksec (in five contiguous exposures)
with the ACIS-I instrument at the focus.  The first, third, and last
of these exposures were taken in subarray mode to reduce the effect of
pileup on bright sources (see Appendix A).  We used the CIAO
software package\footnote{Available at http://asc.harvard.edu/ciao/.}
to perform analysis of the point sources, as described in GHE01 and
Heinke et al.\ 2003 (in preparation).  Two of the brightest sources, 
identified with ROSAT as X5 and X7 (Verbunt \& Hasinger 1998) are
unusually soft in an X-ray color-magnitude diagram (figure 3 in
GHE01) when compared with the population of (optically identified) CVs.
X5 and X7 are also much brighter than the X-ray active binaries (BY Draconis
and RS CVn systems) and radio-identified 
millisecond pulsars identified in the \Chandra\ observation.  X5 and X7
are each less than 1.5 optical core radii 
(r$_c$=23$''$, Howell et al.\ 2000) from the center of 47 Tuc, so we
consider them certain cluster members.  Fits to simple 
blackbody (BB) spectra of X5 and X7 give implied BB radii of 1-2 km. No other
sources with sufficient counts for spectral fitting in the central
region of the cluster showed implied blackbody radii greater than 0.1 km. 

To calculate luminosities and radii, we need estimates of the
distance and $N_H$ column. We recalculate the distance to 47 Tuc using
the mean and standard deviation of the 9 
recent independent distance measurements discussed in Zoccali et al.\ (2001,
fig. 7), weighting all measurements equally.  We find
(m-M)$_V$=13.45$\pm$.11, (m-M)$_0$=13.27$\pm$.13, and thus 
d=4.50$\pm$.27 kpc as our fiducial estimate of 47 Tuc's distance and
1$\sigma$ uncertainty.   The optical extinction to 47 Tuc is
well-known through optical 
studies, which give a mean optical extinction $E(B-V)=0.055\pm.007$
(Gratton et al.\ 1997, 
Zoccali et al.\ 2001).  We calculate a galactic $N_H$ 
column towards 47 Tuc of $N_H=3.0\pm0.4\times10^{20}$ cm$^{-2}$,
using $N_H/E(B-V)=5.5\pm0.4\times10^{21}$ cm$^{-2}$ from
Predehl \& Schmitt (1995), with R=$E(B-V)/A_V$=3.1$\pm0.1$. 

We find absorbed fluxes (0.5-2.5 keV) of $4.3\times10^{-13}$ and
$4.8\times10^{-13}$ ergs cm$^{-2}$ s$^{-1}$ for X5 (excluding eclipses and
dips, see Sect. 2.1) and X7 respectively.
We thus derive intrinsic luminosities for X5 and X7 of $L_X$(0.5-2.5 
keV)=$1.4\times10^{33}$ ergs s$^{-1}$ and $1.9\times10^{33}$ ergs 
s$^{-1}$, and bolometric luminosities of $2.1\times10^{33}$ and
$3.4\times10^{33}$ ergs s$^{-1}$ (from the simplest
hydrogen-atmosphere fits below; we note that these luminosities are
model-dependent). We recalculate (assuming blackbody emission) Verbunt \& 
Hasinger's (1998) ROSAT HRI fluxes\footnote{Using PIMMS at
http://asc.harvard.edu/toolkit/pimms.jsp.} for X5 and X7 to be 
$3.5\times10^{-13}$ ergs cm$^{-2}$ s$^{-1}$ and $4.7\times10^{-13}$
ergs cm$^{-2}$ s$^{-1}$. 
The ROSAT-measured flux of X7 agrees well with the
\Chandra\ luminosity,  while the ROSAT exposures of X5 
probably include dips and eclipses (see sect. 2.1), which may  
explain the 19\% smaller average flux.  We conclude the
observations seem to be consistent with constant luminosity
over long time intervals, and thus with thermal NS emission.

\subsection{Variability}

The time series were analyzed using tools in the IRAF/PROS X-ray
analysis package and the Lomb-Scargle periodogram (Scargle 1982). 
We prepared lightcurves by binning the exposure into 300 parts of 264.2
seconds each, giving an average of $\sim$20 counts per bin.  The
lightcurve of X5 shows significant variability, in particular 
three clear eclipses and significant dips (Figure 1).  The dips near 5
and 10 hours each show $\geq$99\% significance for 
 variability, according to both K-S and Cramer-von Mises tests (Daniel 1990).
The three egresses of the eclipses are 
sharp (less than 30 seconds, see Fig. 1a), whereas the two ingresses
show a dip of roughly 
half the normal flux for $\sim2000$ seconds before the eclipse.
An epoch folding tool in PROS, {\tt period}, gives an 8.72 hour
period, but the near-total absence of flux during the eclipses (see
Fig. 1) allows a simpler period calculation.  We first transform the
event arrival times to the solar system barycenter using the CIAO tool
{\tt axbary}.  By measuring the times between two pairs of eclipse ingresses  
(defined as the last received photons before eclipses), and a pair of
successive eclipse exits (defined as the first received photons after
eclipses), we find a value  of 8.666$\pm$0.005 hours for the
period.  Measuring the eclipse lengths for the second and third eclipses as
the time between the last and first received photons (excluding the
sole photon halfway through the second eclipse) gives 
2488$\pm$12 seconds.  Formal errors are the standard deviations of the
measured time lengths, but are probable underestimates due
to the small number of data points.  The barycentered times of
mid-eclipse (time system TBD) for the second and third eclipses are
MJD=51619.67459 and 51620.03580. 

X7 (the light curve for which is shown in GHE01) shows no significant
variability, 
according to a K-S test.  However, a period search out to a frequency
corresponding to the shortest known LMXB binary period (11 minutes; 4U
1820-30, cf. Stella et al.\ 1987) finds a marginal peak
(false alarm probability [FAP, Scargle 1982]=5.2$\times10^{-2}$) in the power
spectrum at 5.50 hours.  In the
framework of the deep crustal heating hypothesis, any such
period would be most likely due to the changing $N_H$ column, implying
an intermediate inclination angle.  Such a period would be believable for the
type of system expected, but is of low significance.  Our followup {\it
Chandra} observations may confirm or refute it.

The lightcurve of X5 shows eclipses and dips that would not have been visible
in the ROSAT data due to the lower signal-to-noise ratio (but may be
visible in folded light curves; we defer this to later work).  The dips
directly before the eclipses suggest an accretion stream or impact
point obscuring the NS, while the other dips and concomitant 
increases in the $N_H$ column (see Sect. 2.2) are typical of occulting
blobs in the edge of an accretion disk, as seen in dip sources (e.g., Parmar et
al. 1986). Assuming a NS primary of mass 1.4 $M_{\sun}$, and using
Kepler's Third Law, the orbital separation is 

\begin{equation}
a=1.76\times10^{11}\times(0.5+0.5 [M_2/0.3M_{\sun}])^{1/3} {\rm cm} 
\end{equation}
(where $M_2$ is the secondary mass).  We can also use the length of
the eclipse, the orbital separation, and the period of the system to
constrain the secondary radius,

\begin{equation}
 R\geq 3.6\times10^{10} (\frac{1.21\times M_1}{M_1+M_2})(0.5+0.5(M_2/0.3\Msun))^{1/3}{\rm cm}
\end{equation}
Thus, for masses of 1.4 \Msun\ and 0.3 \Msun\ the secondary must have a
radius of at least $3.6\times 10^{10}$ cm (using the Paczynski form
of the Roche lobe radius equation, Frank, King, \& Raine 1992).  The
size of the 
secondary's Roche lobe, determined by the masses and separation of the two
components, must be equal to or larger than the secondary radius.
This allows us to set a minimum secondary mass of 0.17
$M_{\sun}$, for a 90 degree inclination and 1.4 $M_{\sun}$ primary,
regardless of the interior structure of the secondary.  We can also
geometrically constrain the inclination angle.  Using $M_2\leq$0.53 \Msun\
from the \HST\ photometry of Edmonds et al. (2002), and assuming a
spherical secondary, $i\geq$76\fdg1
for a 1.4 \Msun\ primary (or $i\geq$75\fdg6 for a 1.3 \Msun\ primary).  

The size of X5's secondary and the photometry of Edmonds et al. (2002)
rules out a helium white dwarf, and indicates that the
secondary is likely to be a main-sequence star.  The main-sequence
nature of the secondary implies the 
accreting material is hydrogen-rich.  Rapid settling of heavy elements  
onto the NS surface means that for very low accretion rates (such that the
observed X-ray luminosity is primarily from deep crustal heating 
as in BBR98) the upper layers of the atmosphere
should be essentially pure hydrogen (Romani 1987).  Therefore we focus
on hydrogen atmosphere 
models for X5.  The possible period of 5.5 hours for X7 also suggests
a main-sequence secondary, and a hydrogen atmosphere.  We note that
X5's eclipsing and dipping (with increased $N_H$ column) behavior is
very similar to that of the quiescent 
neutron stars 4U 2129+47, as reported by Nowak et al. (2002), and MXB
1659-29 (Wijnands et al., 2002), although X5's dipping
behavior seems to be less regular.

\subsection{Spectra}

To fit the spectra of X5 and X7 in detail, we prepared the event files without
removing cosmic ray afterglows, as this tends  
to remove valid events in the cores of bright sources (\Chandra\ ACIS team
advice\footnote{Available at
http://www.astro.psu.edu/xray/acis/recipes/}).   This
change increased the total flux by $<$10\%.     We used the $\sim$2\arcsec
regions produced by WAVDETECT to extract most ($\gtrsim94$\% for a $\leq$1.5
keV source) of the counts for each source, and took background spectra
from nearby areas without bright sources. We excluded times in which
X5 suffered eclipses and dips from our initial spectral analysis.  Background
and cosmic-ray contamination remains negligible, with only
$\sim$20 background events (estimated) 
compared to 4181 and 5508 total events in the X5 and X7 source
extraction regions.  We extracted the spectra in
205 energy bins, each 73 eV wide, giving a spectrum that is optimally 
oversampled by a factor of two, and grouped the spectral bins to
$\geq$20 counts/bin (to ensure applicability of the $\chi^2$
statistic) for all but the highest energy bin.   Below 0.5 keV the  
energy calibration remains uncertain, and we have few counts
($<2$\%), so we fit the spectra from 0.5 to 10 keV, yielding 31 bins
for both X5 and X7.  The recent tool for correcting the ancillary
response function for the time-dependent degradation of the ACIS
low-energy QE degradation\footnote{See
http://cxc.harvard.edu/cal/Acis/Cal\_prods/qeDeg/.}
has been used in all calculations. The primary effect of correcting
for this degradation is a $\sim$30-40\% decrease in the required neutral
hydrogen column, 
with the other parameters relatively unaffected.  It seems
unlikely that further ACIS calibration refinements will greatly affect the
results we present here.
 
Calculations
with the CIAO PIMMS software of the observed flux showed that both
sources should suffer pileup: approximately 9\% of X5's counts
(allowing for eclipses) and 10\% of X7's counts are expected to be
recorded in an ACIS pixel during the 3.2 sec ACIS integration time
when another photon also lands there.  Pileup is a 
complicated problem which has not yet been thoroughly understood.  We
use the Davis (2001) ACIS pileup model, implemented in ISIS\footnote{ISIS is
available at http://space.mit.edu/CXC/ISIS/.}(Houck \& DeNicola, 
2000); see Appendix A for details of the pileup analysis.  

We fit the spectra of X5 and X7 in XSPEC, using blackbody,
Raymond-Smith, thermal bremsstrahlung, 
power law, and neutron star atmosphere models,
each including 
photoelectric absorption (as a free parameter) and the Davis pileup
model. 
We have used the unmagnetized hydrogen and helium neutron star
atmosphere models of Lloyd, Hernquist, \& Heyl (2002),  and the unmagnetized
hydrogen and solar-metallicity neutron star atmosphere models of
G\"ansicke, Braje, \& Romani (2002)\footnote{Available at
http://heasarc.gsfc.nasa.gov/docs/xanadu/xspec/models/gbr.html} both with 
temperature, radius, and (gravitational) redshift as free parameters.
(Redshift has not been allowed as a free parameter in previous studies
of qLMXBs due to poor count statistics, but these spectra allow
self-consistent constraints in redshift and radius together. See
Sect. 3.2.)  The
atmospheric plasma (assumed to be completely ionized) affects the 
spectra through coherent electron scattering and free-free
absorption. The models of Lloyd et al. (2002) incorporate an improved
treatment of the Gaunt factors, and give similar results as the
G\"ansicke models within a few percent.  We use both models to give an idea
of the dependence of our conclusions upon model differences.
We used the surface gravity ($g_{s}$) of a canonical NS,
$\log{g_{s}}=14.38$ (appropriate for a 1.4 $M_{\sun}$, 10 km NS) although two
alternative surface gravity models of D. Lloyd where
$\log{g_{s}}=14.20$ 
or 14.50 were also tested. The differences were found to be minor, as
expected (Romani 1987, Zavlin et al. 1996), with no significant
parameters affected by more than a few percent.  
Magnetic fields up to $10^{11}$ G should
not affect these fits in either flux or $T_{eff}$ (Lloyd et al.
2002).  We fix the
distance at 4.5 kpc; our estimated distance uncertainty (see Sect. 2) is 6\%,
and affects both M and R linearly.  Any 
 distance error will be the same for both sources, so for comparison
 we do not include distance uncertainty in Sect. 3.2 and the
figures. We allow the pileup parameter $\alpha$ to vary 
freely, and note that it remains between 0.45 and 0.55, as expected
from current pileup testing (see App. A).

 With the pileup convolution, several single-component
models (blackbody, 
thermal bremsstrahlung, power law, and the NS atmosphere models, all
with photoelectric absorption as a free parameter) fit the
spectra of X5 and X7 reasonably well (null hypothesis prob. $>5$\%).
No Raymond-Smith
thermal plasma model with abundances above 1\% solar fit the data
acceptably.  Single-component power-law  
models require a spectral photon index $>5$, which would be highly unusual
(pulsars with power-law X-ray spectra generally display indices $\sim$1-3).
Thermal bremsstrahlung models require kT below 600 eV, which is much softer
than in cataclysmic variables (CVs) that display high X-ray luminosity
(Eracleous et al.\ 1991), and the $F_X/F_V$ ratios are higher than
those of known CVs.  Although we cannot formally exclude the blackbody
models, we note difficulties with this model: the derived radii
are rather small (while the lack of X-ray pulsations in other qLMXBs
suggests that the emission is isotropic, Chandler \& Rutledge 2000),
and recent accretion is
expected to leave an atmosphere of H or He on the NS.  We
include parameters for blackbody fits in Table 1 for comparison with
previous studies.  
Helium atmospheres are ruled out for X5 due to the main sequence nature of the
companion (Edmonds et al. 2001).  We note that tests of D. Lloyd's He
atmosphere models forced 
much higher redshifts for standard radii than obtained below for both
neutron stars, leading
to inconsistency with the causality constraint.
The hydrogen atmosphere models give temperatures of order $10^6$ K,
and the best 
fits imply masses and radii in the range expected of neutron stars,
leading to our focus upon these models.   Using solar
metallicity photoelectric absorption, the fits for X5 give
$\chi^2_{\nu}$=1.38 with 26 degrees of freedom (dof), for a null
hypothesis probability (prob)=10\%. For X7, we obtain 
$\chi^2_{\nu}$=1.20, 26 dof, prob=22\%.  The spectra of X5 and 
X7 along with the Lloyd H-atmosphere model predictions and the fit residuals
are plotted in Figure 2.    

  The photoelectric
absorption ($N_H=3.0\pm0.4\times10^{20}$ cm$^{-2}$) derived from
optical studies (see Section 2) is ruled out for both sources at
 90\% confidence.  For X7, $N_{H,22}>0.09$ ($N_{H,22}$=$N_H$/$10^{22}$
cm$^{-2}$) is required, while $N_{H,22}>0.04$ for X5.  
We fit separately the intervals of X5's spectrum which showed dips
(count rate $<13$ cts bin$^{-1}$, see Fig. 1), and those
which did not.  The dipping portions (with 19\% of the total
counts) showed $N_{H,22}$, 0.24$^{+.08}_{-.14}$  (compared to
$N_{H,22}=0.09\pm0.05$ outside dips) when $T_{eff}$ and z are held equal
to the best values outside dips.  We use the times outside dips for
all other fits in this paper.

 We applied the best-fitting hydrogen atmosphere spectral fits from the full
datasets to the subarray exposures of  
 X7 and X5. Freezing all parameters except the pileup parameter
$\alpha$, we obtained $\chi^2_{\nu}$/dof=1.36/11, prob=18\% for X5,
and  $\chi^2_{\nu}$/dof=0.76/11, prob=68\% for X7.  These fits to the full
datasets are shown with the 
subarray data in Fig. 3.  The agreement between the subarray and full
dataset parameters demonstrates that the pileup formalism is valid.  We
use the complete datasets (minus X5's eclipses and dips) for the
remainder of this paper, as the subarray fits are poorly
 constrained due to the lack of counts.

\subsection{Possible Edges and Power-Law Constraints}

We then performed spectral fits adding photoelectric absorption of gas with
the nominal metallicity of 47 Tuc 
(20\% solar for iron-group, 40\% solar for $\alpha$-group elements,
taken to be calcium and below, Zoccali et al.\ 2001) to the known
galactic absorption column.  However, the fits become
poorer for X7 as metallicity in the excess gas decreases, due
to a negative residual near 0.7 keV.  X5 produces
$\chi^2_{\nu}$/dof=1.21/31, null hypothesis probability (prob.) of 8\%, and X7
produces $\chi^2_{\nu}$/dof=1.41/26, prob. 11\%.  Adding 
an absorption edge improves the residuals for both X5 and X7 (see
Fig. 4), and increases the null hypothesis prob. values to 18\% and
83\%.  The edge  
energy is best fit at 660$^{+45}_{-67}$ eV for X5 and
632$^{+36}_{-52}$ eV for X7 (90\% conf. intervals).  The F-test
calculator in XSPEC gives 9.7\% and
0.02\% as the respective probabilities that X5 and X7 do not
require the additional component (but see Protassov et al.\ 2002
 for limitations of this statistic).  This makes the feature in X5
marginal at best, but the similarity to X7 is very suggestive.
The solar metallicity photoelectric absorption fits are similarly
improved by the addition of the edge.  Other sources in 47 Tuc (CVs
with more complex spectra; Heinke et al. 2003 in prep) do not
seem to show a similar edge, arguing against an instrumental
interpretaion.  Nevertheless, we also continue to 
consider the simpler hydrogen atmosphere with solar metallicity
absorption model, for comparison with results from other systems.  We
show results for spectra using 47 Tuc metallicity gas and an edge with 
parameters free in Table 1, along with results using solar metallicity
gas, and a blackbody with 47 Tuc metallicity gas and edge. 

Subtracting a gaussian (modeling an absorption line) instead of
fitting an edge also improves the fits to $\chi^2_{\nu}$=0.98 for X5, and 
$\chi^2_{\nu}$=0.75 for X7.  The gaussian line energies are best
fit at 0.74$^{+.03}_{-.03}$ keV and 0.71$^{+.03}_{-.04}$ keV for X5 and X7
respectively, with $\sigma<90$ eV for both.  This gives a better fit
to X5 than the 
edge model.  However, since hydrogen-atmosphere incandescent neutron stars
should not have cyclotron lines (as their fields should be too low),
nor any atomic transition lines (such as discussed in Sanwal et
al. 2002), the subtraction of gaussians is not physically motivated
for a pure hydrogen atmosphere in the deep crustal heating model.  If
the features are indeed atomic 
lines, then the atmosphere is not pure hydrogen and the models will not
accurately describe the NS radii, as discussed in Section 3.2.  We
discuss the possibilities for the edge in Section 3.1.

The qLMXBs Cen X-4 and Aql X-1 have generally shown a hard power-law
 tail, of photon index 1-2, which predominates over the thermal
 component above 2 keV.  Such a tail is 
 absent from X5 and X7, where the energy received above 2 keV can
 be attributed wholly to pileup.  The subarray exposures show only 3
 or 4 counts (out of 369 or 423 counts for X5 and X7 respectively)
 beyond 2.5 keV.  An F-test of the 
 confidence for addition of a hard powerlaw component with photon
 index of 1.5 gives F=1.56 for X5 and F=0.2 for X7,  
 indicating the component is not required (probability of achieving these
 F-values by chance if the component is not real is 22 and 66\%
 respectively). In 
 Figure 5, the best hydrogen-atmosphere fits to X5 and X7 with an
 added PL of index 1.5 are shown, in an XSPEC $\nu F_{\nu}$ plot for ease of
 comparison with other results.  The upper limit to power law
 component flux in the 0.5-10 keV band is only 0.6\% of the total
 flux for X7, and 4\% for X5.    

\section{Discussion}

\subsection{Edge possibilities}

The possible edge identified in the spectra of X5 and X7, if
unredshifted, is most likely identified primarily with the 
O\textsc{V} edge, at 0.627 keV.  This edge would suggest 
gas at temperature T$\sim3\times10^4$ K  (Kallman \& McCray 1982),
which we speculate may be a wind from the disks.  Assuming  a wind of
density structure $\rho=\rho_0 R^{-2}$, the ionization parameter of
the wind $\xi=L/nR^2$ remains constant until the wind is 
slowed by interactions with the cluster medium as the wind's density
approaches the cluster gas density ($n_e=0.067$ cm$^{-3}$,
Freire et al.\ 2001) at $R_{out}\sim 10^{16}$ cm.  The mass
loss rate in the wind outflow can then be calculated, using 

\begin{equation}
\dot{M}=\frac{M_{tot}}{\delta t_{Rin-Rout}}=\frac{4 \pi \rho_0 R_{out}}{R_{out}/v} 
\end{equation}

\noindent
where 
$M_{tot}$ is the total gas mass in the correct ionization state
(considered to be the wind density structure out to $R_{out}$), and
$v$ is the gas velocity.   

  Using 40\% solar abundance for oxygen in
47 Tuc (Zoccali et al.\ 2001), and a K-shell photoionization
cross-section for OV of
$\sigma$=0.47$\times10^{-18}$ cm$^2$ (Daltabuit \& Cox 1972), we find
that the optical depth $\tau=\kappa \int \rho(R) dR= 40 \rho_0
R_{in}^{-1}$.  If we choose $R_{in}=10^{10}$ cm (which
is 1/4 of the circularization radius of the disk) and
optical depths from Table 1, we derive $\xi\sim10$, which is sufficient
to keep O in the OV state.  A wind extending from $R=10^{10}$ out to
$3\times10^{16}$ cm reproduces the edges' optical depth, giving a total
ionized wind mass of $\sim10^{-9}$\Msun\ for both systems. 
We can also derive 

\begin{equation}
\dot{M}=8.1\times10^5 \tau R_{in}{\rm \, g \, s}^{-1}=
1.3\times10^{-20} \tau R_{in} {\rm \, \Msun \, yr}^{-1}
\end{equation}

\noindent
by mass conservation (from eq. 3) and using the gas sound speed 
$v=\sqrt{3 k T/m_p}=2.7\times10^6$ cm s$^{-1}$.  
This leads to a mass loss rate
of order $5\times10^{-11} \Msun$/yr.  This is significantly more than
the highest allowed (by its quiescent luminosity) current low-level accretion
rate of the neutron star 
($2\times10^{-13}$ \Msun\ yr$^{-1}$).  This suggests that
most of the material in the accretion disk is swept out by winds
instead of ever reaching the neutron star (see Sect. 3.3).  We note
that this mass loss rate is of the same
order as the time-averaged accretion rate necessary to heat the
neutron stars to an incandescent luminosity of a few $10^{33}$ ergs
s$^{-1}$ (BBR98; see Introduction).

This
calculation demonstrates the plausibility of this model, but of course
other scenarios cannot be ruled out.  Another obvious possibility is
that the edge represents the prominent 0.9 keV OVIII photoionization
edge (blended 
with other metal lines; BBR98) in neutron star atmospheres with nonzero
metallicity, gravitationally redshifted by a factor (1+z)
of 1.35 to 1.43 (reasonable for canonical NSs).  This inspired us to
attempt to fit a solar-metallicity neutron star atmosphere model,
produced by G\"ansicke, Braje, \& Romani (2002), with absorbing $N_H$
columns for either solar-metallicity 
or 47 Tuc-metallicity absorption.  These models were not successful, generating
$\chi_{\nu}^2=1.3$ with 32 dof for X5 (requiring $N_H>2\times10^{21}$)
and $\chi_{\nu}^2=2.3$ with 
29 dof for X7.  However, subsolar metallicity neutron star atmospheres
(implied by the known metallicity of 47 Tuc) have not yet been
calculated, and may fit these data.    

\subsection{Constraints on mass and radius}

 The ``radiation radius'' $R_{\infty}$, as seen by an observer at
 distance D, is calculated by measuring the observed flux and
 temperature.  Using $F_{\infty}=\sigma
 (R_{\infty}/D)^2 T_{\infty}^4$ and $T_{eff}=T_{\infty} (1+z)$ (e.g.,
 Lattimer \& Prakash 2001), where
 z, the gravitational redshift of a star with mass M and true radius
 R, is given by  $z=-1+1/\sqrt{1-2GM/Rc^2}$, we get  

\begin{equation}
R_{\infty}=\frac{D\times(1+z)^2}{T_{eff}^2}\sqrt{\frac{F_{\infty}}{\sigma}}  
\end{equation}
 
\noindent
 The true radius $R$ is recovered with $R=R_{\infty}/(1+z)$. 
The normalization of the Lloyd models is conveniently expressed
as norm=(R [km]/D[10 kpc])$^2$, 
while that of the 
G\"ansicke models is expressed as d[pc]/R[10 km] = 1/$\sqrt{{\rm
norm}}$, or R[km] = 10 d[pc] 
$\sqrt{{\rm norm}}$.  Thus the true radius can be straightforwardly
calculated from the normalization parameter alone.
 The mass can then be derived using $M=R c^2 (1-(1+z)^{-2})/(2 G)$.
 
The surface gravity $g_s=GM(1+z)/R^2$ is used to calculate the opacity
of the hydrogen atmosphere.  The sense of alterations in the surface
gravity is that a lower surface gravity model gives a slightly cooler
temperature, and thus requires a lower redshift to fit the data than a
higher surface gravity model.  This is a relatively small effect due
to increased pressure ionization with gravity (Romani 1987), and
alters mass and radius estimates by less than 10\% in the region of
parameter space of interest.  In this section we use models with 
log($g_s$)=14.38, appropriate for a 1.4 \Msun, 10 km NS.   
  
The X-ray color magnitude diagram for 47 Tuc (Figure 2 in GHE01) shows
that while X7 is brighter than X5, it is also significantly softer.
This effect cannot be attributed solely to a hard PL component in X5,
nor to pileup 
(pileup hardens the raw spectrum, and the brighter X7 has slightly
more pileup), nor to differences in $N_H$ column.  The spectral difference is
confirmed by the  
significant difference in the best-fit blackbody temperature between
X5 and X7 (see Table 1).  If one holds  z 
constant at 0.306 (appropriate for M=1.4\Msun\ and R=10 km), the lower
temperature for X7 would require a significantly 
larger radius.  However, an effect of increasing the
gravitational redshift is to lower the apparent temperature of the NS
surface, and thus to bring the two spectral fits to similar radii.
The assumption of the relation z=0.306 does not encompass the full 
range of neutron star EOS predictions even if M=1.4\Msun. 
Thus we allow redshift to
vary in our fitting procedure, and explore the space in M and R that
is allowed by the constraints on z and normalization. 

The 90\% confidence contours in M-R space are plotted for
X5 (red dotted) and X7 (blue dashed) using simple solar-metallicity
photoelectric 
absorption and the Lloyd et al. (2002) hydrogen-atmosphere model,
in figure 6a.  The same contours are shown in figure 6b for the
G\"ansicke hydrogen atmosphere model.  In figure 7, the contours for
X5 and X7 are shown using 
absorbing gas set to 47 Tuc metallicity, with an absorption edge
(7a uses the Lloyd model, 7b uses the G\"ansicke model).  A
sample EOS, the APR model (Akmal et al.\ 1998) is also 
plotted, as is the causality constraint (above which the required sound speed
exceeds the speed of light).  The M-R space above the causality line cannot 
contain NSs under any EOS.   

X5 is well-fit by a 1.4 \Msun\ 10 km NS, while X7 is not consistent
with masses below 1.8 \Msun\ for typical NS predicted radii between 9
and 14 km, using either the solar metallicity absorption
fits or 47 Tuc metallicity absorption fits with an edge.  Some
spectral fits (fig. 7a) require X7 to have a mass exceeding 3 \Msun\ at
90\% confidence.  
For a standard radius of 10 km, we  
have calculated the 90\% mass range for both X5 and X7 in Table
1.  We have also calculated the radius range for an assumed redshift of
0.306; the large radii required for X7 exclude the possibility of a
strongly heated polar cap.   We note that the models of G\"ansicke et 
al. (2002) slightly decrease the large mass or radius of X7, since they give
slightly higher temperatures.  The difference between the G\"ansicke
and Lloyd models is small, but leads to significant
differences in the mass and radius ranges.  We give the 90\% mass and
radius ranges 
as above for the G\"ansicke models also (see Table 1) as an indication of the
range in models, and plot the 90\% contours
derived from both the Lloyd and G\"ansicke models in figures 6 and 7.
Lloyd model fits to X7 
with 47 Tuc metallicity gas and no edge (less likely due to the poorer
quality of these fits, see Sect. 2.2) also require a mass of
2.32$^{+.15}_{-.19}$ \Msun\ for a 10 km radius.  Similar G\"ansicke model fits
allow a mass of 2.11$^{+.16}_{-.45}$ \Msun.  Lloyd model fits with a
gaussian absorption line 
require a mass of 2.29$^{+.14}_{-.05}$ \Msun, while equivalent G\"ansicke model
fits require a mass of 2.18$^{+.14}_{-.02}$ \Msun\ for the same 10 km 
radius assumption.  {\it Thus none of our spectral fits allow for X7
to have the canonical NS mass and radius.}

The implication that X7 is more massive than the canonical NS range
(1.3-1.45, Thorsett \& Chakrabarty 1999) is very interesting.  Two
possibilities 
for the increased mass present themselves.  X7 may have been born more
massive than most NSs (see e.g. Timmes, Woosley \& Weaver 1996), or it may
have accreted significant mass from its companions over its long life.
Observations of either type of massive NS are more probable in 
globular clusters compared to the field.  Initially massive NSs
($>1.8$ \Msun) cannot 
find their way into relativistic NS-NS binaries (which have the best mass
determinations) because such systems require a common-envelope stage,
and massive neutron stars probably descend from stars over 20 \Msun\ 
which do not pass through a red giant stage (see Barziv et al. 2001).
Alternatively, a NS in a dense globular cluster has many chances to accrete
mass from its numerous companions over its long life (see Hurley \&
Shara 2002, Grindlay et al. 2002), possibly leading to massive NSs as
inferred in some LMXBs (e.g. Orosz and Kuulkers 1999).  The
objection that observed MSPs, probable descendants of LMXBs, are not
more massive than 1.4 \Msun\ (Thorsett \& Chakrabarty 1999) can be
answered if accretion of 
substantial ($>0.1$ \Msun) mass buries the magnetic field of the NS.
In either case, confirmation of a NS more massive than 1.8 \Msun\ would
immediately exclude many NS equations of state (see Lattimer \& Prakash, 2001).
  
  However, we note a possible problem with the possibility that X7 is 
more massive 
than 1.8 \Msun.  Its continued brightness over thirty years in which it
was not observed to be actively accreting suggests that it does not experience
enhanced URCA cooling, while some other NSs may require direct URCA
cooling to explain their low luminosity (Cen X-4, Colpi et al. 2001;
SAX J1808.4-3658,  Campana et al. 2002).   More
massive neutron stars are expected to cool faster due to the higher
density in their cores (see Kaminker et al.\ 2001), so one would
expect X7 to be dimmer than most qLMXBs if its accretion history is
similar.  Of course, X7 could spend a larger fraction of its history
accreting at high rates, explaining the apparently high core
temperature.  Or most LMXBs could be as massive as X7 (as implied by
some QPO results, van der Klis 2000).

Alternatively, both X7's apparently high mass and/or radius and the
$\sim$0.64 keV possible absorption features in both X5 and X7 may be  
explained if X5 and X7 derive a substantial portion of their
X-ray luminosity from continued low-level accretion.  Accretion at rates
sufficient to produce $L_X\sim10^{33}$ ergs s$^{-1}$ would also be
sufficient to maintain metals in the photosphere, altering the
opacities slightly closer to a blackbody and leaving a pronounced O
VIII edge (combined with other metal edges) at
0.87/(1+z) keV.  Although solar-metallicity NS atmospheres do not fit
the data well, subsolar-metallicity NS atmospheres (not yet
calculated) may. Such models would
predict smaller radii and/or masses for X7 than our
hydrogen-atmosphere fits, perhaps consistent with canonical NS
predictions.  These models would need to 
self-consistently calculate the effects of accretion at low rates on the
emerging NS spectrum, and would suffer from uncertainties in the
metallicity of accreting matter.   

\subsection{Power law and accretion}

One of the major results of this study is the result that the
quiescent LMXBs X5 and X7 show very little or no hard power-law spectral  
component, unlike most field qLMXBs.  A lack of evidence for a PL
 tail has also been noted in the qLMXBs CXOU 132619.7-472910.8 
 in $\omega$ Cen (Rutledge et al.\ 2002a) and U24 in NGC 6397 (Grindlay et
 al. 2001b), but X5 and X7 allow much tighter limits on the PL
component. Although the origin of the PL component is unclear,
no explanation from deep crustal heating has been given, so 
it is generally assumed to be due to continued ADAF-like 
low-level accretion (e.g., Campana 1998a, Rutledge et 
al. 2001b, 2002a).  We note that the spectra and luminosities of black
hole accretion disks in quiescence are
similar to the PL component alone in qLMXBs containing NSs (power laws with
spectral index $\alpha\sim$1-2 and $L_X\sim10^{30-33}$; Kong et 
al. 2002), perhaps implying a similar accretion-related origin for both.  
   Within the deep crustal heating explanation for the thermal 
emission, the lack of a power law component and lack of variability
 in X5 and X7 suggest that accretion onto the NS surface is held at a very low
level or completely stopped.   
Low binary mass transfer rates compared to Cen X-4 and Aql X-1 could
explain the relatively weak power law and lack of outbursts in X5 and
X7.  However, if the deep crustal heating model 
is correct, the current X-ray luminosity will depend upon the long-term
($\lsim$10$^{4-5}$ years) accretion history of the NS.  The low mass
transfer explanation fails to account for the higher luminosity of X5 and 
X7, $L_X({\rm 0.5-10 keV})\sim10^{33.2}$, compared to
$\geq$10$^{32.2}$ for the more active Cen X-4 (Rutledge et al.\ 2001a).    
And accretion to the disk is clearly 
continuing in X5, as shown by the dips and blue optical color of the
companion (Edmonds et al. 2002). 

 It is possible that 
little to no mass from the disk is reaching the neutron star, 
but is instead being driven from the system in a wind, as in the ADIOS
solutions for black hole accretion (e.g., Blandford \& Begelman 1999,
and see Menou \& McClintock 2001).
The relative lack of
accretion onto the neutron star surface could be explained by a
propeller effect (when the 
magnetospheric radius expands 
beyond the corotation radius and expels inflowing material; Illarionov
\& Sunyaev 1975) or a pulsar wind.  The shock
generated by a propeller effect, or the shock from a
pulsar wind at lower accretion rates, could also provide an 
explanation for the hard power 
law X-ray emission (see Campana \& Stella 2000). The
transition to a propeller effect in the expected luminosity range
($L_X\sim10^{36}$ ergs s$^{-1}$) may have been observed in Aql X-1 (Zhang et
al.\ 1998, Campana et al. 1998b; but cf. Maccarone \& Coppi 2002, and 
Chandler \& Rutledge 2000)
and 4U 0115+63 (Campana et al.\ 2001).  A pulsar wind is
thought to eject matter falling from the Roche-lobe overflowing
secondary in the millisecond pulsar PSR J1740-5340 in NGC 6397 (Burderi
et al.\ 2002).   The advent of a pulsar wind may 
sweep away the entire accretion disk due to the radial 
dependencies of the pulsar radiation pressure and the disk pressure
(Burderi et al.\ 2002).  However, this assumes a standard Shakura-Sunyaev disk
structure throughout the entire disk. Due to dynamical encounters in
globular clusters (cf. Grindlay et al. 2002), it may be possible to cycle
between pulsar emission and accretion regimes.

We note a possible correlation between the existence of a strong power law
component and recent (suggesting frequent?) outbursts.  Cen X-4, Aql
X-1, CX1 in NGC 6440 (in't Zand et al.\ 2001), KS 1731-260 (Wijnands
et al. 2001), and 4U 2129+47 (Nowak et al. 2002)
each have strong PL components and recorded (recent) outbursts, while
X5 and X7, U24 in 
NGC 6397 (Grindlay et al.\ 2001b), and CXOU 132619.7-472910.8 in
$\omega$ Cen (Rutledge et al.\ 2002a) have exhibited 
evidence of neither (although the limits on the 6397 and $\omega$ Cen
sources' PL components are currently weak).  This may indicate a
difference in the mode or level of accretion activity between the two groups.

\section{Conclusions}

The X-ray sources X5 and X7 are thermally
radiating neutron stars with hydrogen atmospheres, probably heated by transient
accretion.  X5 shows eclipses, which allow parameters of
the binary system to be inferred.  Both X5 and X7 are well-fit by the
hydrogen atmosphere model spectra of G\"ansicke et al. (2002) and
Lloyd et al. (2002), absorbed by a column of gas with 
the known cluster metallicity and displaying a possible absorption
feature near 0.64 keV tentatively 
identified with an OV edge.  The feature may instead be
intrinsic to the neutron star atmosphere, in which case it is most
likely identified with the 0.87 keV (rest-frame) complex of oxygen and
metal lines in a subsolar-metallicity neutron star atmosphere.   A hard
power-law component to the spectra, such 
as has been observed in other qLMXBs, is extremely weak or nonexistent in
X5 and X7.   The well-known distance to 47
Tucanae allows modeling of the neutron star atmospheres to constrain
a space in mass and radius for each, if the atmospheres are purely
hydrogen.  Fits using the ISIS pileup 
model and hydrogen atmosphere fits require X7 to be more 
massive than 1.8 \Msun\ for the range of radii 9-16 km, while the
constraints on X5 are consistent with a mass of 1.4 \Msun\ for a wide range
of radii.   However, if the majority of the X-ray luminosity is
derived from low-level accretion, then appreciable amounts of metals
would remain in the neutron 
star atmospheres.  In addition to producing the
feature near 0.64 keV, the metals would alter the overall shape of the
spectrum slightly closer to a blackbody, possibly allowing both X5 and X7
to fit a canonical NS mass and radius.  If accretion is not
continuing, a high mass for X7 is unavoidable.

  A series of 
observations ($\sim$300 ksec total) in October 2002 with {\it Chandra}, using
the back-illuminated ACIS-S chips for 
greater sensitivity below 4 keV, will give us more data with which to
 constrain the mass and radius, and mode of emission, of these neutron
stars.  We will be particularly interested in looking for intrinsic
variability,  and confirming or refuting the possible edge feature.
The eclipsing  
behavior of X5 also allows the possibility of using \HST\ to measure the radial
velocity of the secondary, constraining the mass of X5.  Unfortunately 
optical spectroscopy of X7 is beyond the reach of \HST, due to crowding
 (Edmonds et al.\ 2002).  These demonstrations of fits with
hydrogen atmosphere models show that the project of 
constraining the neutron star equation of state through spectral
fitting holds particular interest for these two systems.

\acknowledgments

This work was supported in part by \Chandra\ grants GO0-1098A and GO2-3059A.
C.H. thanks M. Nowak, J.~E. McClintock, J. Raymond, J. McDowell, F. Walter,
P. Kondratko, and J. Lattimer for helpful discussions.  
C.H. also thanks M.~C. Miller for EOS relations, the anonymous referee
for helpful suggestions, B. G\"ansicke and
K. Arnaud for assistance with XSPEC implementation of NS models, and
especially J. Houck 
for assistance with ISIS.  We thank the \Chandra\ ACIS team at Penn
State and the CXC team at the CfA for advice on data analysis.

\appendix

\section{Treatment of Pileup Effects}

Since X5 and X7 possess soft spectra, cut off 
below 1 keV by the falling detector response and neutral hydrogen
absorption, their raw spectra can be modeled to zeroth order as
gaussians with peaks at energy $\sim1$ keV and widths of
$\sigma\sim0.3$ keV from a simple fit to the pulse height data.  Thus, 
the pileup will produce a secondary peak at $\sim2$ keV with $\sigma$
$\sim0.4$ keV, containing 9\% and 10\% for X5 and X7 respectively of the total
received flux, neglecting
grade migration. (Grade migration occurs when the charge clouds from
two photons are recorded 
at the same time in adjacent pixels, and the pattern of charge is
assigned an inferior grade, leading to the discarding of the event;
see Davis 2001).  This secondary peak was misinterpreted in GHE01 (where 
pileup was ignored) as a high energy tail.  We 
added a gaussian component to model the pileup component of the
spectra, allowed the parameters to 
vary freely, and discovered that the best-fitting gaussian component
gave a median energy of 2.23 or 2.25 keV for X5 and X7, $\sigma$=0.38
keV for both, and a total flux fraction (considering that the
effective area at 1.1 and 2.2 keV differs by a factor of 3/5)
of 9\% and 10\%, respectively.  The addition of this gaussian improved the
reduced chi-squared statistic
by a factor of 2-4 for any single or multiple-component spectral model we
tried, and the close agreement of the empirical parameters with the
expected ones confirms this simple pileup model.  We note that such a
simple model is only fortuitously valid for X5 and X7 since they lack
significant hard spectral components, as discussed in Sect. 2.3.

A more complete treatment of pileup is offered in the Davis (2001)
model of ACIS pileup\footnote{For discussion of pileup analysis, see
http://space.mit.edu/CXC/analysis/davis/head2002/index.html}.  This
has been implemented separately in XSPEC (Arnaud et al.\ 1996) 
v. 11.1, and the MIT spectral analysis system ISIS.  Comparison of the
two systems suggested that, at the time of our 
analysis, the ISIS implementation produced more accurate values for
the source model parameters\footnote{see
http://space.mit.edu/CXC/analysis/PILECOMP/index.html for discussion},
so we used the ISIS 
system for our spectral analysis. The Davis pileup model parametrizes grade 
migration by assuming that the good grade fraction of piled-up events is
proportional to $\alpha^{(p-1)}$, where p is the number of photons in
one event detection cell during one frame time.  We used a piled psf
fraction (psffrac) of 0.95, and fixed $g_0$ (the good grade branching
parameter) at 1.0 following Davis (2001), 
while the parameter $\alpha$ was allowed to vary. In effect this alters
the normalization of the pileup bump for this low-pileup exposure.  
Davis (2001) found
that $\alpha=0.5\pm0.02$ for a heavily piled-up quasar.  However, this
grade migration parametrization is unlikely to perfectly describe the
function. The parameter $\alpha$ may vary depending on the degree of
pileup and the energy of the photons, therefore on the incident
spectrum, so we do not fix $\alpha$.  For low pileup 
spectra like X5 and X7, most of 
the pileup consists of only two photons. Thus only one parameter is
needed to describe the grade migration for most of the pileup
(effectively a normalization of the primary piled peak), and the
exact form of the parametrization is not important.  We confirm this
by varying the parameter psffrac between 0.9 and 0.95 and letting
other parameters vary.  When psffrac is altered, $\alpha$ also varies
in the opposite direction, but all other source parameters remain
constant to roughly 1\%.  We
find that the best fit value for $\alpha$ generally lies between 0.45
and 0.55 for all models that fit the spectrum accurately ($\chi^2_{\nu}<1.2$).

Our interpretation of the pileup is supported by three small parts
(4.6, 3.2, and 0.8 ksec) 
of the full {\it Chandra} dataset that we intentionally collected using a
subarray mode to reduce the 
frame time to 0.9 s, and thus reduce pileup in any bright sources.
ISIS fits suggest that the pileup fraction is 7 and 8\% respectively for
the full datasets of X5 and X7, but only 2\% for the subarray
exposures.  The difference can be clearly
seen between figure 3, the 
subarray exposure spectra of X5 (uneclipsed, 369 counts) and X7 (423
counts), compared to Figure 2 
showing the full combined exposures of X5, excluding eclipses and dips
(4181 counts), and X7 (5508 counts).  The pileup bump beyond 2 keV is mostly
removed in the subarray exposure.

\clearpage


\begin{deluxetable}{ccc}
\tablewidth{3.2truein}
\tablecaption{\textbf{Spectral Model Parameters}}
\tablehead{
\colhead{\textbf{Model Parameter}} & \colhead{\textbf{X5}} &
\colhead{\textbf{X7}} 
}
\startdata
\hline
\multicolumn{3}{c}{\textbf{Blackbody, Z$_{{\rm solar}}$ absorption}} \\ 
\hline
kT, eV & $237^{+7}_{-1}$ & $217^{+5}_{-11}$ \\
$N_{H,22}^a$ & $.03^{+.02}_{-0}$ & $.03^{+.04}_{-0}$ \\
$\chi^2_{\nu}$/dof & 1.37/27 & 1.11/27  \\
Null hyp. prob. & 9.5\% & 31\%\\
PL flux \% & $0^{+4}_{-0}$ & $0^{+0.6}_{-0}$ \\
R$_{\infty}$, km & $1.86^{+.2}_{-.1}$ & $2.43^{+.5}_{-.1}$ \\
\hline
\multicolumn{3}{c}{\textbf{Lloyd H-atm, Z$_{solar}$}} \\
\hline
kT$^b$, eV & $119^{+21}_{-18}$  & $84^{+13}_{-12}$  \\
$N_{H,22}^a$ & $0.09^{+.05}_{-.05}$ & $.13^{+.06}_{-.04}$ \\
$\chi^2_{\nu}$/dof & 1.38/26 & 1.20/26 \\
Null hyp. prob. & 10\%  & 22\%  \\
PL flux \% & $0^{+3}_{-0}$ & $0^{+0.5}_{-0}$ \\
R$^b$, km & $11.7^{+7.5}_{-3.5}$ & $34^{+22}_{-12}$ \\
M$_{NS}^c$ & $1.4^{+0.7}_{-0.7}$  & $2.3\pm0.1$  \\
\hline
\multicolumn{3}{c}{\textbf{G\"ansicke H-atm, Z$_{{\rm solar}}$}} \\
\hline
kT$^b$, eV & $140^{+17}_{-18}$  & $106^{+10}_{-1}$  \\
$\chi^2_{\nu}$/dof & 1.37/26 & 1.22/26 \\
R$^b$, km & $8.2^{+4.0}_{-2.7}$ & $18.6^{+9.6}_{-5.1}$ \\
M$_{NS}^c$ & $0.9^{+0.9}_{-0.9}$  & $2.1^{+0.4}_{-0.3}$  \\
\hline
\multicolumn{3}{c}{\textbf{Lloyd H-atm, Z$_{{\rm 47\, Tuc}}$, edge}} \\
\hline
kT$^b$, eV & $101^{+21}_{-14}$  & $74^{+13}_{-8}$  \\
$N_{H,22}^a$ & $0.09^{+.08}_{-.05}$ & $0.11^{+.09}_{-.06}$ \\
$\chi^2_{\nu}$/dof & 1.25/24 & 0.73/24 \\
Null hyp. prob. & 18\% & 83\% \\
Edge E, eV & $660^{+45}_{-67}$ & $632^{+36}_{-52}$ \\
Max $\tau$ & $0.32^{+.31}_{-.20}$ & $0.56^{+.27}_{-.22}$ \\
PL flux \% & $1.2^{+2.2}_{-1.2}$ & $0^{+0.5}_{-0}$ \\
R$^b$, km & $19.0^{+8.8}_{-7.8}$ & $50^{+21}_{-15}$  \\
M$_{NS}^c$ & $2.0^{+0.4}_{-0.6}$ & $2.7^{+0.2}_{-0.2}$ \\
\hline
\multicolumn{3}{c}{\textbf{G\"ansicke H-atm, Z$_{{\rm 47\, Tuc}}$, edge}} \\
\hline
kT$^b$, eV & $125^{+20}_{-16}$  & $94^{+11}_{-12}$  \\
$\chi^2_{\nu}$/dof & 1.26/24  & 0.74/24  \\
R$^b$, km & $11.5^{+6.1}_{-4.0}$ & $27.2^{+10.6}_{-9.2}$  \\
M$_{NS}^c$ & $2.0^{+0.4}_{-1.2}$ & $2.6^{+0.3}_{-0.4}$ \\
\enddata
\tablecomments{ All errors are 90\% confidence limits.  Distance of
4.50 kpc is assumed.  Galactic column of 3$\times10^{20}$ is imposed
as minimum $N_H$.  
$^a$ $N_{H,22}$ in units of 10$^{22}$ cm$^{-2}$.  $^b$  
R, kT for assumed z of 0.306, implying 10 km, 1.4 \Msun\ NS; this
tests for consistency with the standard model.   $^c$  Range of
M$_{NS}$ derived from allowed z values if radius fixed at 10 km. 
}
\end{deluxetable}

\clearpage


\vspace*{0.3cm}
\psfig{file=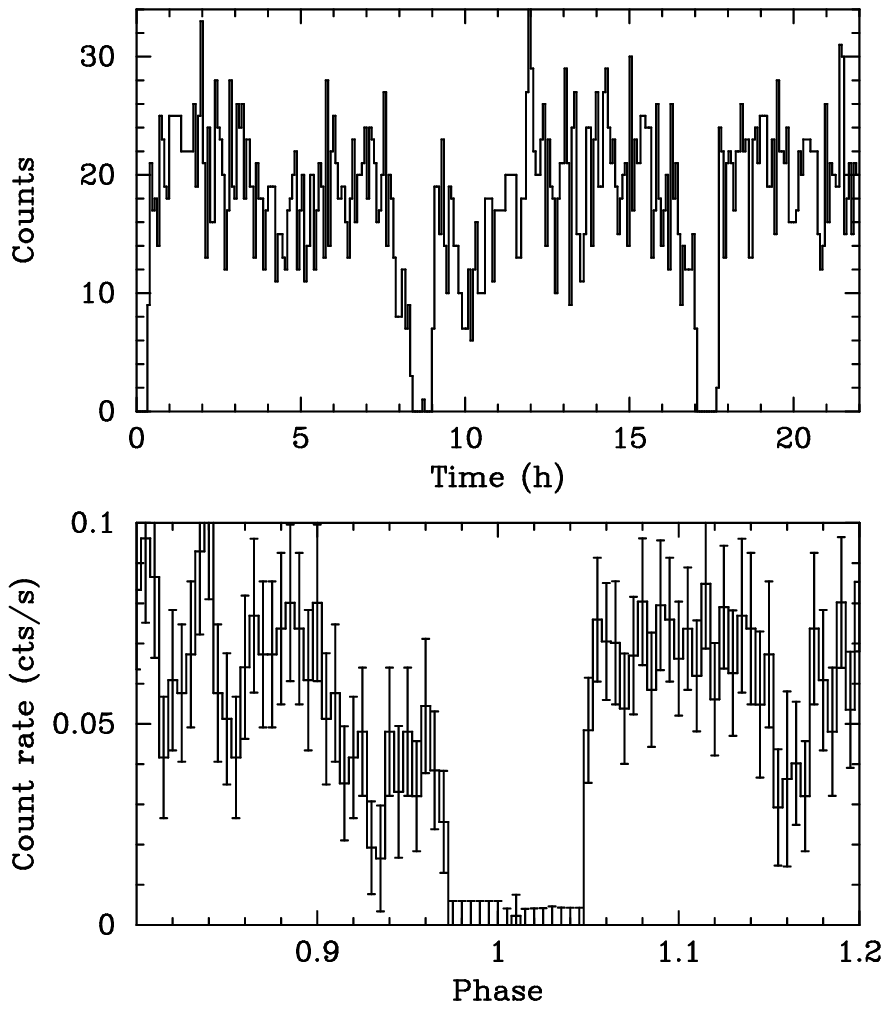}

\vspace{0cm}

\figcaption[f1.eps]{Top: Lightcurve of X5, showing the three
eclipses and other dips. Bins are each 264
seconds long, and the
small gaps between exposures have been removed.  Bottom: Phase plot
of the eclipse 
and pre-eclipse dip portion of X5's lightcurve, using 8.666 hour
period. Bins are 156 s long. \label{Figure 1}} 

\psfig{file=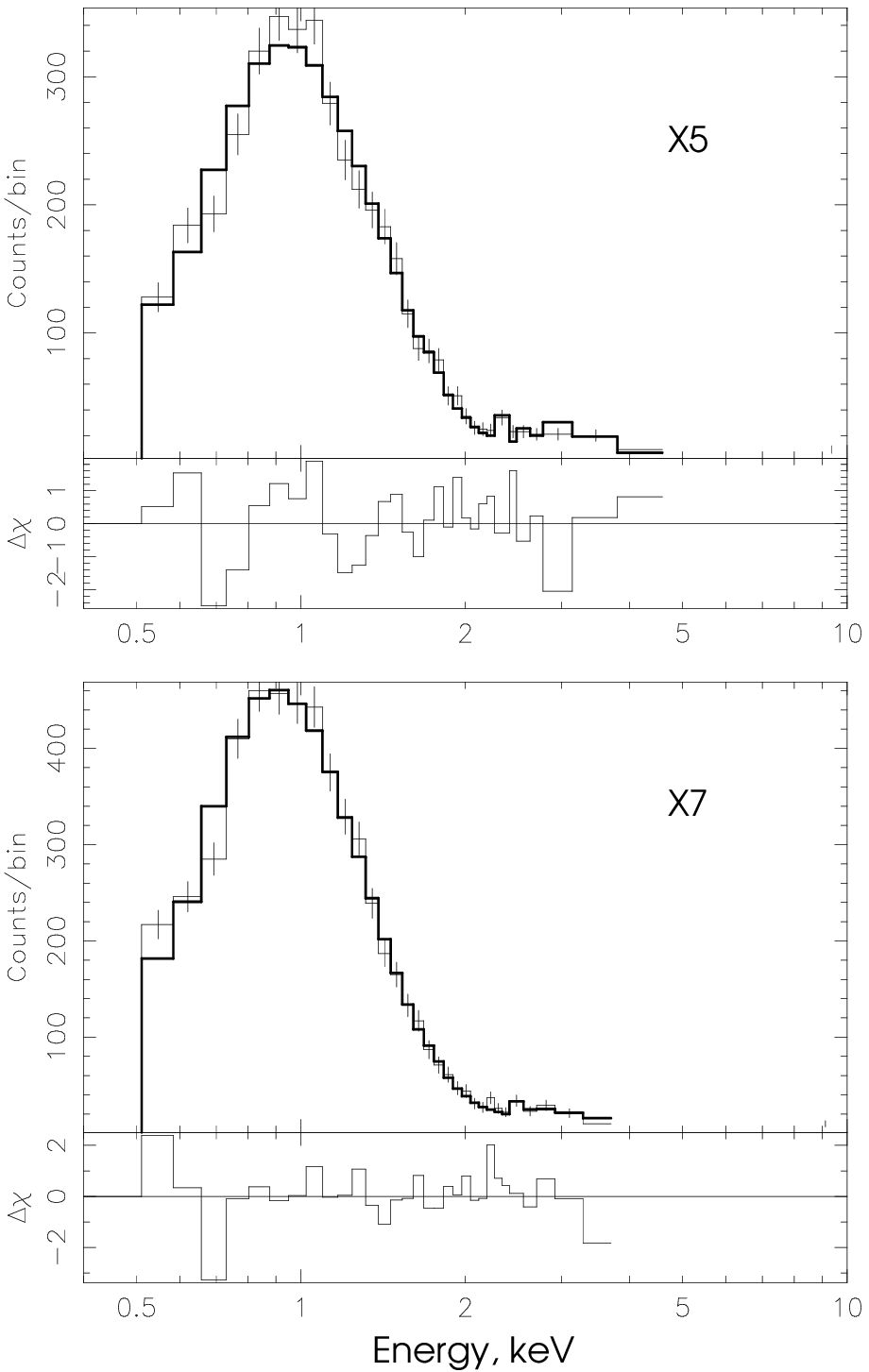}

\figcaption[f2.eps]{Data and best-fit Lloyd model spectra
using hydrogen atmosphere model, solar metallicity gas for X5 and X7.
Note the residual feature near 0.7 keV in both spectra, and lack of
a harder component. \label{Figure 2}} 

\psfig{file=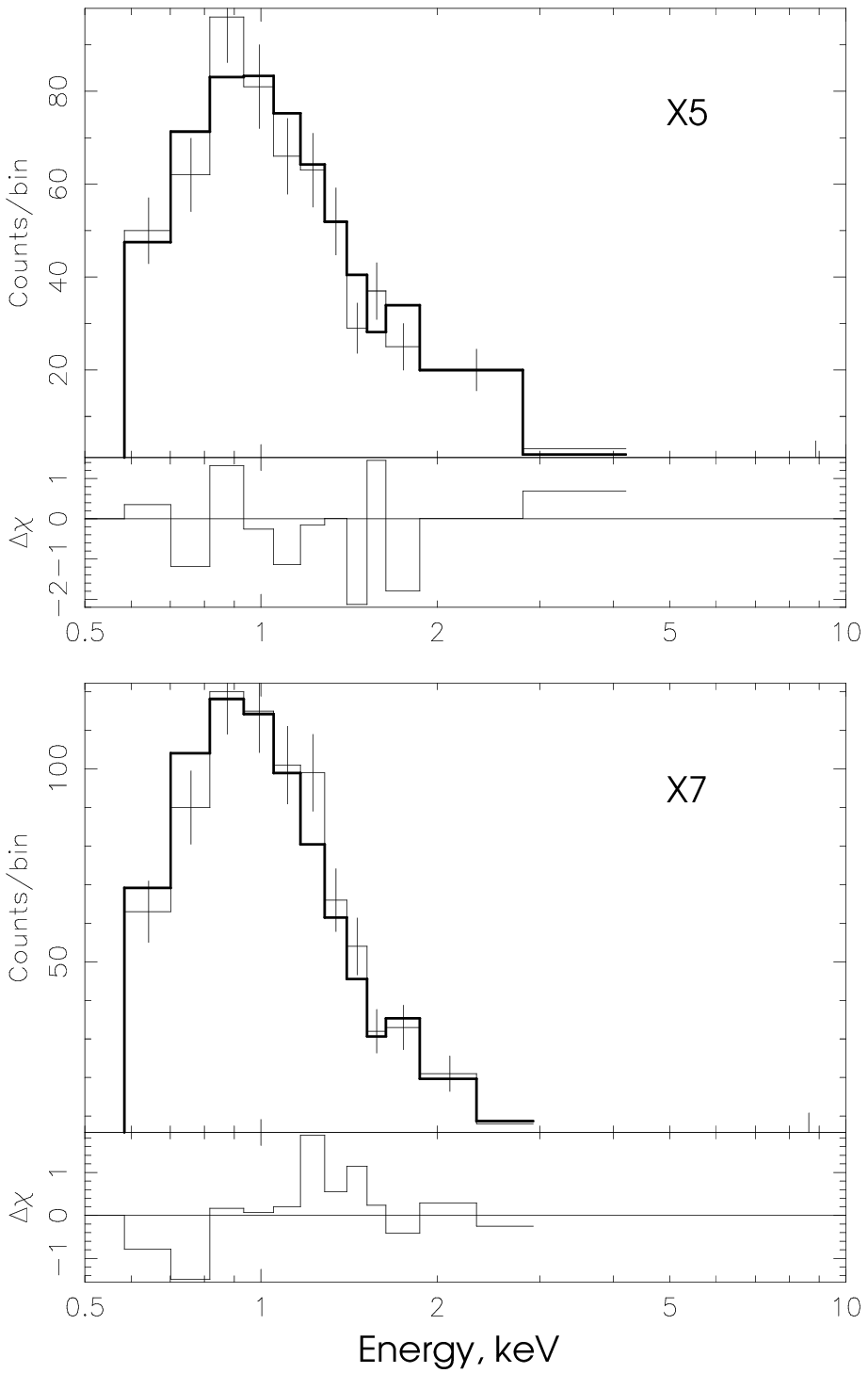}

\figcaption[f3.eps]{Best-fit Lloyd hydrogen atmosphere model to
full datasets (as in Fig. 2), plotted over the subarray exposure data
 for X5 and X7.  Note diminishment of pileup bump (compared to Fig. 2)
beyond 2 keV due to smaller frame exposure times.  \label{Figure 3}}

\psfig{file=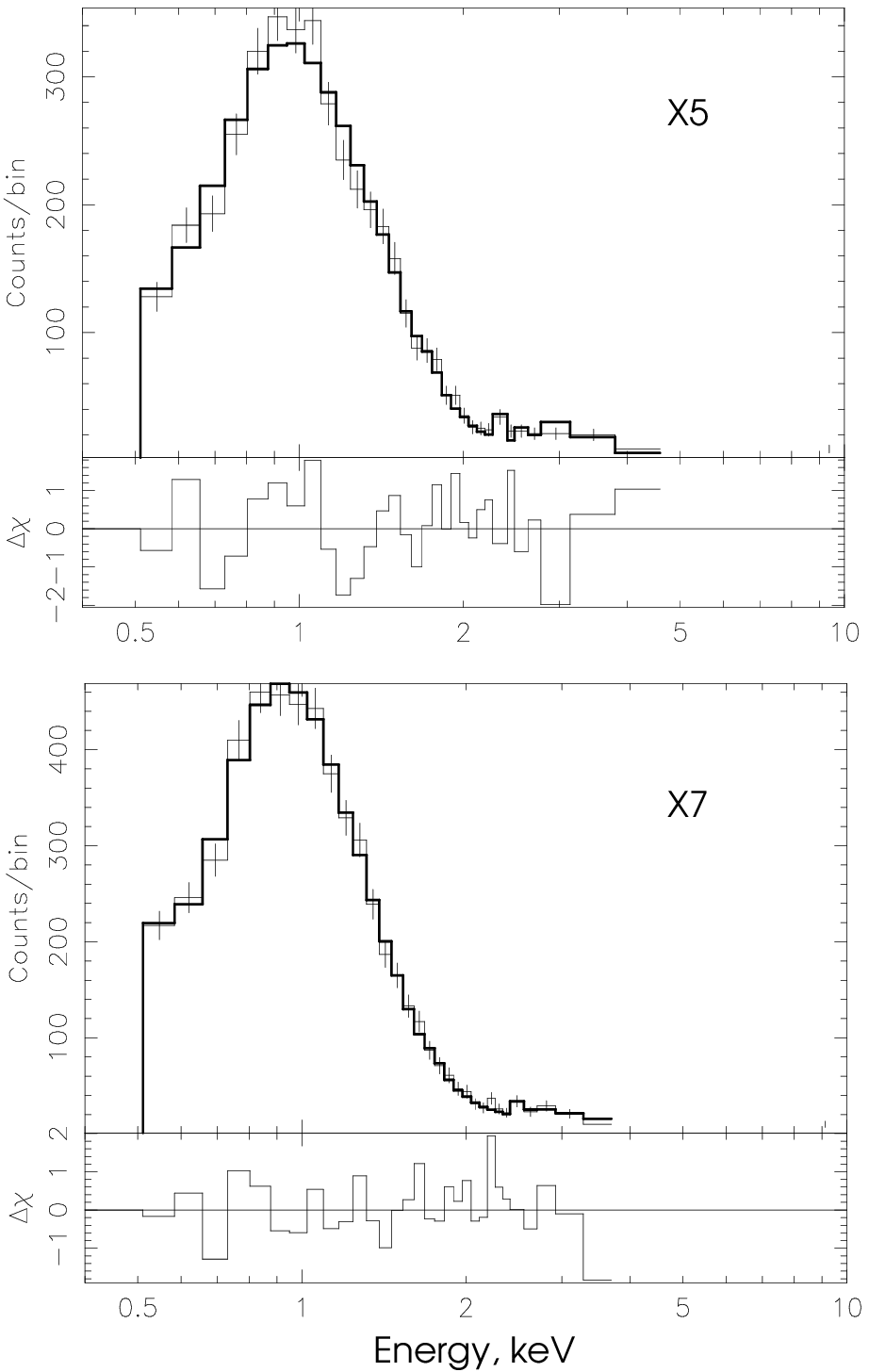}

\figcaption[f4.eps]{Data and best-fit model spectra for X5 and X7
using Lloyd hydrogen atmosphere model, 47 Tuc metallicity gas,
and best-fitting edge.  Note improvement in low-energy
residuals for both X5 and X7 over Fig. 2. \label{Figure 4}}

\psfig{file=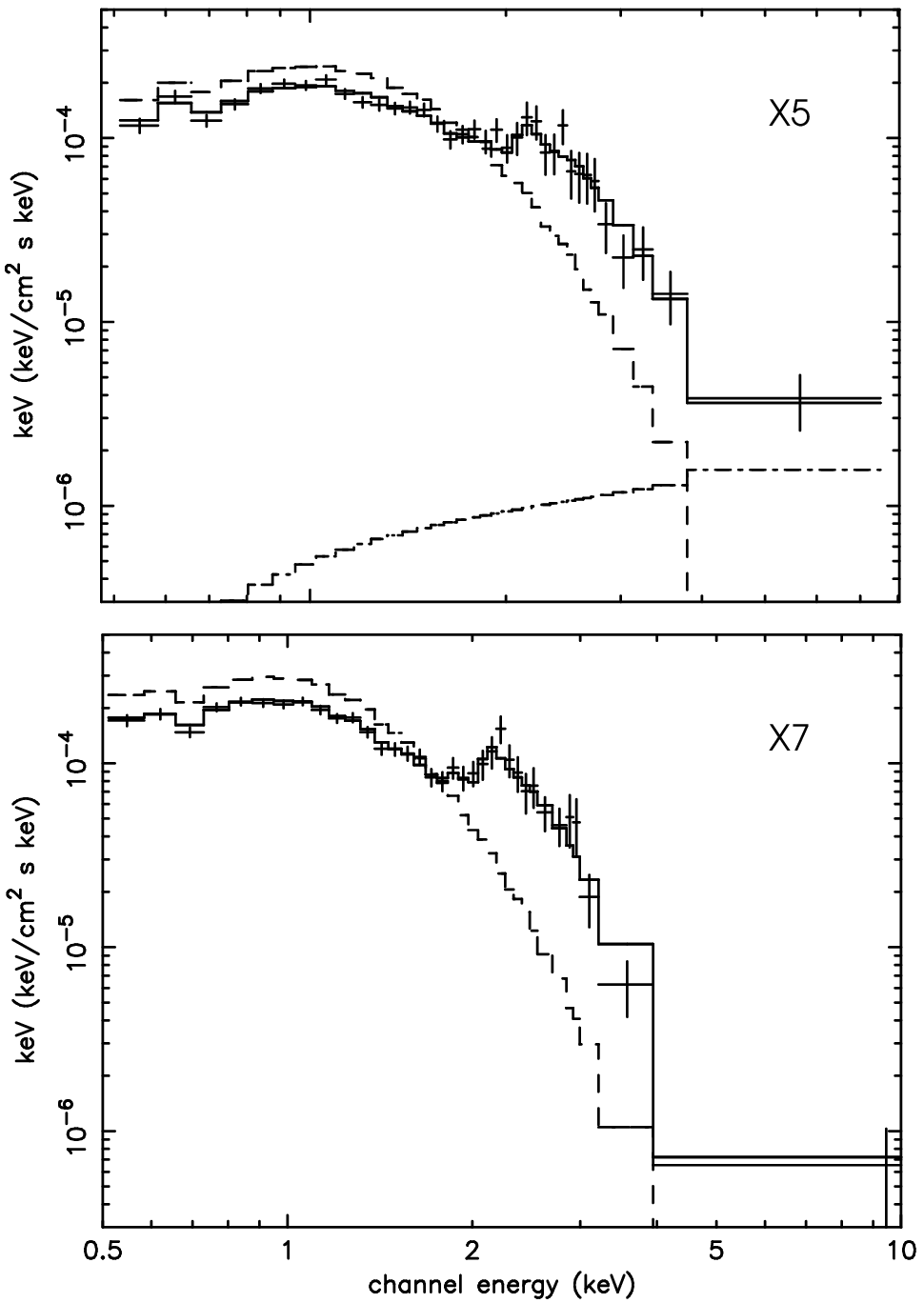}

\figcaption[f5.eps]{Data and model spectra (best-fit Lloyd hydrogen
atmosphere with 47 Tuc metallicity gas, edge, PL), plotted in $\nu  
F_{\nu}$ space, of X5 and X7.  The bump near 2 keV is an instrumental
effect due to pileup, not unfolded but accounted for in the
fit. The two components--the thermal hydrogen NS atmosphere, and the
power law component--are displayed at their best-fit normalizations,
which for X7 is not visible in this plot. \label{Figure 5}}   

\clearpage

\psfig{file=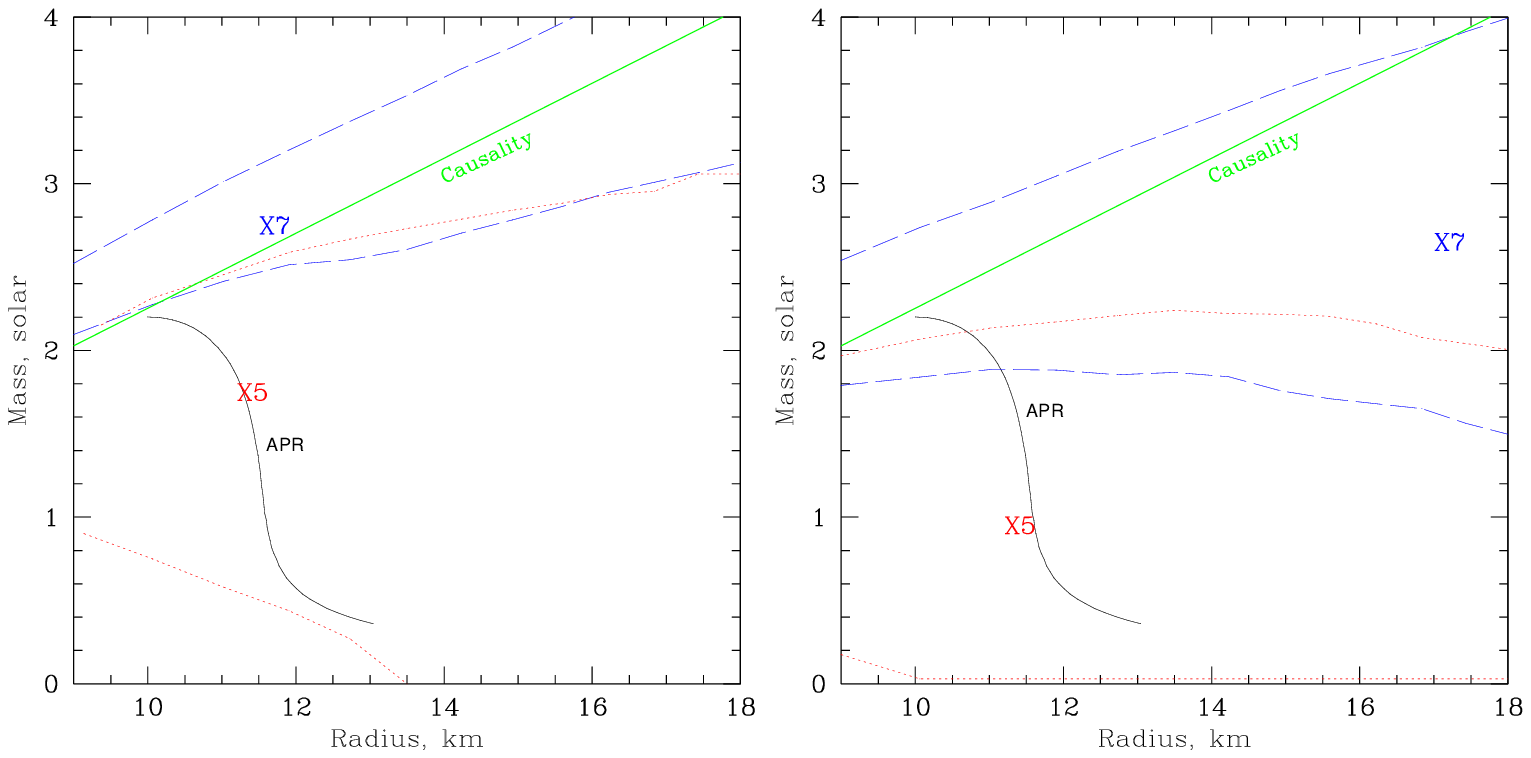}

\figcaption[f6.eps]{90\% confidence contours in the M-R
plane for X7, (blue) long-dashed line, and X5, (red) dotted line, using
the best-fit 
model with hydrogen atmosphere NS model, and standard photoelectric
absorption.  The left panel uses the Lloyd et al. (2002) hydrogen
atmosphere model, while the right panel uses the G\"ansicke et
al. (2002) hydrogen atmosphere model.  The labels ``X5'' and ``X7''
indicate the rough locations 
of the best fits.  The causality constraint is plotted as the 
(green) solid line, and the APR model is plotted as the black
curve. {\it (See the electronic edition of ApJ for a color version of this
figure.)}\label{Figure 6}}

\psfig{file=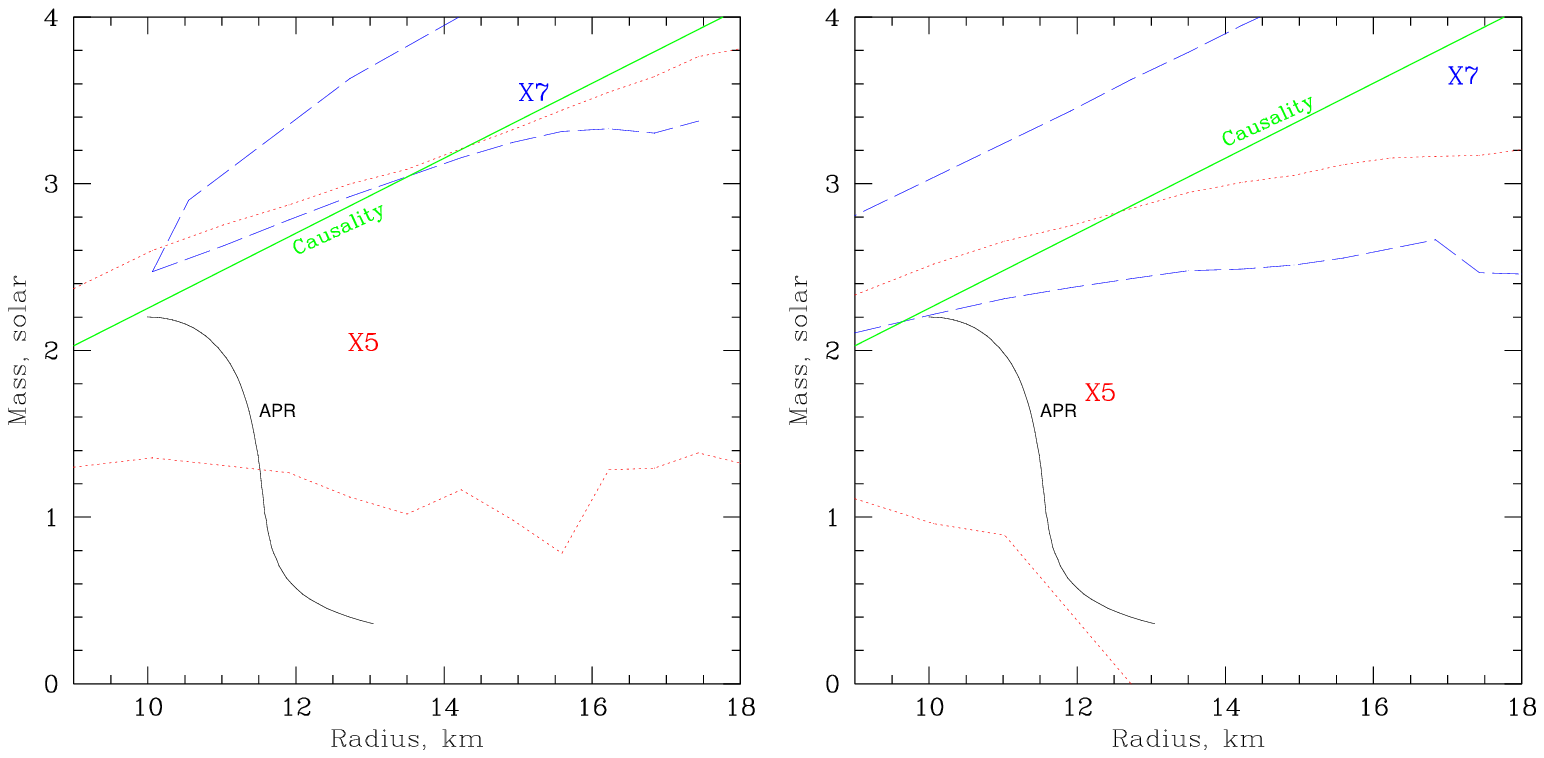}

\figcaption[f7.eps]{90\% confidence contours in the M-R plane for
X7, (blue) long-dashed line, and X5, (red) dotted line, using the best-fit
model with 47 Tuc metallicity absorbing gas, and a free edge. The left
panel uses the Lloyd et al. (2002) hydrogen 
atmosphere model, while the right panel uses the G\"ansicke et
al. (2002) hydrogen atmosphere model.  The
labels ``X5'' and ``X7'' indicate the rough locations 
of the best fits.  The causality constraint is plotted as the (green)
solid line, and the APR model is plotted as the black curve.   {\it
(See the electronic edition of ApJ for a 
color version of this figure.)} \label {Figure 7}}

\end{document}